\documentclass[12pt, onecolumn]{IEEEtran}

\usepackage{cite}
\usepackage{amsmath,amssymb,amsfonts}
\usepackage{algorithmic}
\usepackage{textcomp}
\def\BibTeX{{\rm B\kern-.05em{\sc i\kern-.025em b}\kern-.08em
    T\kern-.1667em\lower.7ex\hbox{E}\kern-.125emX}}

\usepackage[pdftex]{graphicx} 
\usepackage{subfig}
\usepackage{stfloats}
\usepackage{here}
\usepackage{multirow}
\usepackage{tabularx}
\usepackage{bm}

\usepackage{bm}

\title{A numerical exploration of signal detector arrangement in a spin-wave reservoir computing device}

\author{
\normalsize{
{{Takehiro~Ichimura}$^{1}$},
{{Ryosho~Nakane}$^{1}$}, 
{{Gouhei~Tanaka}$^{2}$},
 {{and}}
{{Akira~Hirose}$^{1}$} \\
}
\thanks{This work was supported in part by the JSPS KAKENHI Grant 18H04105, in part by the Cooperative Research Project Program of Research Institute of Electrical Communication (RIEC), Tohoku University, and in part by the New Energy and Industrial Technology Development Organization (NEDO) as a project JPNP16007.}
\thanks{$^{1}$Department of Electrical Engineering and Information Systems, The University of Tokyo, Tokyo 113-8656, Japan (e-mail: ichimura@eis.t.u-tokyo.ac.jp,nakane@cryst.t.u-tokyo.ac.jp,ahirose@ee.t.u-tokyo.ac.jp)}
\thanks{$^{2}$International Research Center for Neurointelligence (IRCN), The University of Tokyo, Tokyo 113-0033, Japan (e-mail: gtanaka@g.ecc.u-tokyo.ac.jp)}
}

\begin{document}

\maketitle

\begin{abstract}
%
This paper studies numerically how the signal detector arrangement influences the performance of reservoir computing using spin waves excited in a ferrimagnetic garnet film. This investigation is essentially important since the input information is not only conveyed but also transformed by the spin waves into high-dimensional information space when the waves propagate in the film in a spatially distributed manner. This spatiotemporal dynamics realizes a rich reservoir-computational functionality. First, we simulate spin waves in a rectangular garnet film with two input electrodes to obtain spatial distributions of the reservoir states in response to input signals, which are represented as spin vectors and used for a machine-learning waveform classification task. The detected reservoir states are combined through readout connection weights to generate a final output. We visualize the spatial distribution of the weights after training to discuss the number and positions of the output electrodes by arranging them at grid points, equiangularly circular points or at random. We evaluate the classification accuracy by changing the number of the output electrodes, and find that a high accuracy ($>$ 90\%) is achieved with only several tens of output electrodes regardless of grid, circular or random arrangement. These results suggest that the spin waves possess sufficiently complex and rich dynamics for this type of tasks. Then we investigate in which area useful information is distributed more by arranging the electrodes locally on the chip. Finally, we show that this device has generalization ability for input wave-signal frequency in a certain frequency range. These results will lead to practical design of spin-wave reservoir devices for low-power intelligent computing in the near future.
\end{abstract}
%

%
\IEEEpeerreviewmaketitle

%
%
%
%



\section{Introduction}


Reservoir computing is a machine learning framework for temporal pattern processing \cite{lukovsevivcius2009reservoir}. This computational framework, generalized from special kinds of recurrent neural networks \cite{esn,lsm,VERSTRAETEN2007391}, has an advantage in its low training cost. A reservoir computing system normally consists of a dynamic reservoir and a readout. The reservoir transforms input time series data into a high-dimensional feature space, and then the readout performs a pattern analysis of the high-dimensional spatiotemporal signals. 

The connection weights in the reservoir are fixed, and only the parameters in the readout are trained with a simple learning algorithm. The computational cost for the readout training is much smaller than those of standard algorithms for recurrent neural networks \cite{jaeger2002tutorial}. Therefore, reservoir computing is a highly promising machine learning framework capable of high-speed learning, which is desirable in particular for sequential data processing in fluctuating environments. The reservoir computing methods have been widely studied for many applications including dynamic system approximation, time series forecasting, temporal pattern classification, and anomaly detection \cite{TANAKA2019100}.


Another attractive aspect of the reservoir computing framework is that its hardware can be realized by a variety of physical phenomena \cite{TANAKA2019100}. Physical reservoir computing attracts much attention these days. The reservoir can be implemented by using electronic devices, photonic and optical devices, mechanical devices, spintronic devices, evolving materials, biological substrates, and many others \cite{TANAKA2019100,van2017advances,brunner2019photonic,dale2017reservoir}. These devices need to satisfy some requirements for achieving the role of a reservoir, such as nonlinearity, high-dimensionality, and history-dependent property. Some studies aim to develop energy-efficient reservoir computing chips for near-future artificial intelligence technology. In particular, spintronic reservoirs are novel candidates for realizing small-size reservoir chip devices \cite{sto,Nakane2018IEEEAccess:_Reser_Compu_with_spin_waves_Excit_in_a_Garne_film,Prychynenko2018,Nomura_2019,Tsunegi2020}.

Recently, we have proposed an on-chip spin-wave-based reservoir computing device that utilizes the nonlinearity and history-dependent property of spin waves propagating through a continuous ferrimagnetic garnet film \cite{Nakane2018IEEEAccess:_Reser_Compu_with_spin_waves_Excit_in_a_Garne_film}. We have also numerically demonstrated a machine-learning estimation of a characteristic included in input-signal waveforms. 

One of the remarkable device features is the fact that the dimensionality of the reservoir output signals can be controlled by adjusting the number of output electrodes (signal detectors) even if the chip size is fixed. Another feature is that the proposed device is promising for realizing energy-efficient on-chip computation as next-generation machine-learning hardware due to the wireless signal transmission in the reservoir \cite{Katayama2016IEEENanotech:_wave_based_neuro_compu_frame_for_brain_like_energ_effic_and_integ}. 

The former feature poses a problem of how we appropriately set the number of output electrodes such that we can extract computationally essential information from the propagating spin waves. In our previous work \cite{Nakane2018IEEEAccess:_Reser_Compu_with_spin_waves_Excit_in_a_Garne_film}, we prepared only a small number of output electrodes with fixed positions. However, the number and positions should influence the properties of the reservoir output signals, computational performance of the device, and finally the total system functionality. Since the information fed to the spin-wave reservoir is transformed into high-dimensional space through the dynamics of the spin waves propagating in the chip, investigating the influence of the output-electrode arrangement is essential to appropriate extraction of the information required for high performance reservoir computing. That is, it is significant to consider how many detectors should be used and how they should be arranged in space.


In this paper, we numerically investigate the effects of the number and arrangement of output electrodes on the computational capability of the spin-wave-based reservoir computing chip. Numerical analyses of complex spin-wave dynamics are aimed at understanding the device characteristics including the relationship between its parameters and the performance. For this purpose, we simulate dynamical waveforms of spin waves in a rectangular garnet film to show spatial distributions of the reservoir states in response to input signals, which are represented as spin vectors and used for a machine-learning wave classification task. We place two input electrodes at two corners of the chip, and arrange the output electrodes at all the simulation mesh pixels having readout connections in the first experiment. After a learning process, we visualize the distribution of the weights of the readout connections, which gives insights into how to design the number and positions of output electrodes. Then, we study how classification performance is influenced by the changes of the number and arrangement of the output electrodes.


\begin{figure}
\centering
\includegraphics[width=0.4\hsize]{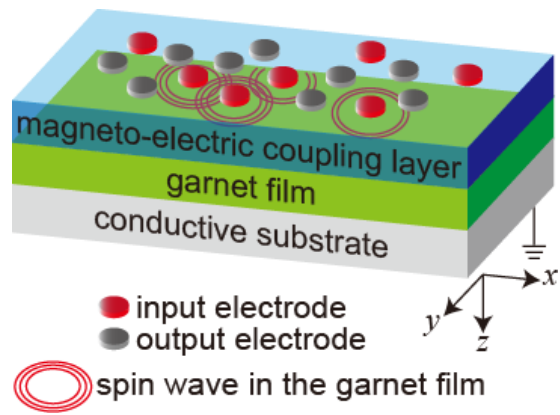} 
\caption{Basic structure of spin-wave reservoir computing device\cite{Nakane2018IEEEAccess:_Reser_Compu_with_spin_waves_Excit_in_a_Garne_film}. (Figure reproduced under permission from IEEE.)}
\label{fig:basicstrct}
\end{figure}

\begin{figure}
\centering
\includegraphics[width=0.7\hsize]{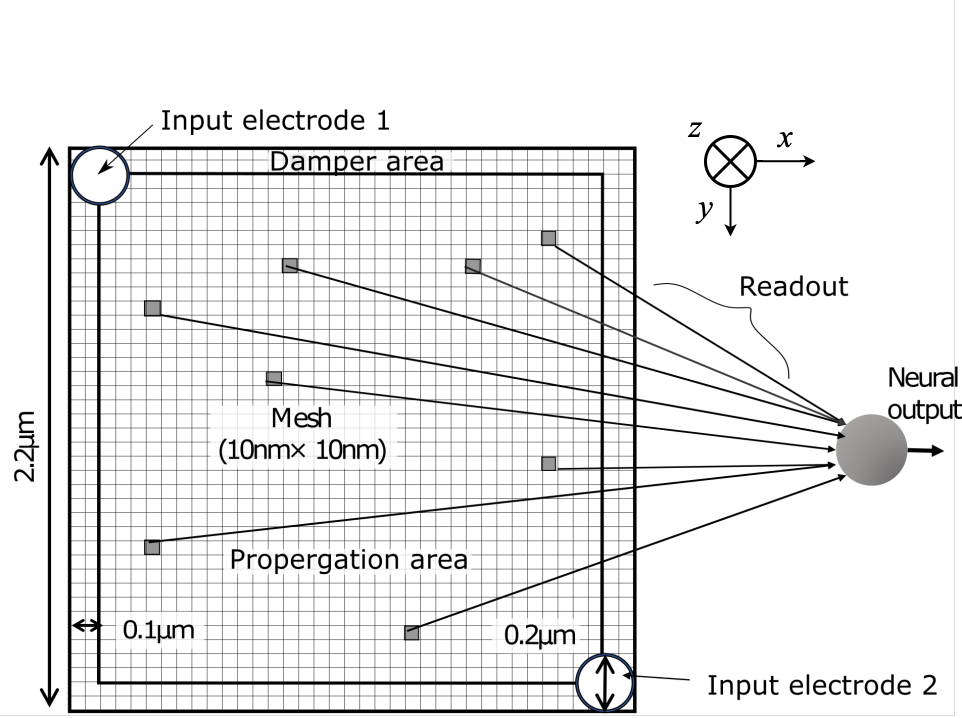} 
\caption{Reservoir chip with two input electrodes (1 and 2 at the top left and bottom right) and the calculation mesh as well as its readout connections.}
\label{fig:sond}
\end{figure}

This paper is organized as follows. Section~\ref{sec:constandtask} describes the basic construction of the spin-wave reservoir and the waveform classification task. In Section~\ref{sec:result}, we show the spin-wave propagation and information distribution. We investigate the influence of the number and arrangement of the output electrodes on the computational performance. We also check the generalization ability in the frequency domain. Section~\ref{sec:conclusion} concludes this paper.

\section{The spin-wave reservoir structure and the task}
\label{sec:constandtask}

\subsection{Physical structure of the spin-wave reservoir}
\label{subsec:const}

Fig.~\ref{fig:basicstrct} shows the structure of the spin-wave-based reservoir computing device proposed in our previous papers ~\cite{Nakane2018IEEEAccess:_Reser_Compu_with_spin_waves_Excit_in_a_Garne_film,Nakane2018ICM:_demon_of_spin_wave_based_reser_compu_for_next_gener_machi_learn_devic}, where spin waves propagate through the ferrimagnetic garnet film. The basic operation is sketched as follows: spin waves are excited by input signals via the input electrodes, they propagate and interfere with one another, and the resultant waveforms are detected by the output electrodes. Since nonlinear phenomena are included in the spin-wave dynamics, this device works effectively as a reservoir, i.e., it transforms a time-series input signal to high-dimensional spatiotemporal signals \cite{Nakane2018IEEEAccess:_Reser_Compu_with_spin_waves_Excit_in_a_Garne_film}.


Fig.~\ref{fig:sond} shows the top view of a garnet film used for spin-wave simulation with MuMax3 \cite{mumax3}, where the entire in-plane area is a square with $2.2 \times 2.2$ $\mu\textup{m}^2$, the vertical thickness is 100 nm, the in-plane mesh size is a square with $10 \times 10$ $\textup{nm}^2$, the vertical mesh length is $50$ nm, the two input electrodes of $\rm 0.2 \mu m$ diameter are located at the top-left and bottom-right corners. In the first numerical experiment below, every mesh unit has a readout connection. The application of input data at the input electrodes changes the uniaxial magnetic anisotropy constant $K_{\rm U}$ along the vertical axis proportionally between the maximum value $K^{\rm H}_{\rm U}$ and the minimum value $K^{\rm L}_{\rm U}$ in the input electrode regions, while it is a constant $K^{\rm H}_{\rm U}$ in other regions. The inner region with $2.0 \times 2.0$ $\mu\textup{m}^2$ has a low damping constant $\alpha$ = 0.001 where spin waves propagate with little decay, whereas the outer frame with 0.1 $\mu$m width has a large damping constant $\alpha$ = 1 where spin waves are significantly damped. It should be noted that the in-plane size is considerably smaller than that ($12 \times 12$ $\mu\textup{m}^2$) in our previous papers~\cite{Nakane2018IEEEAccess:_Reser_Compu_with_spin_waves_Excit_in_a_Garne_film} \cite{Nakane2018ICM:_demon_of_spin_wave_based_reser_compu_for_next_gener_machi_learn_devic} in order to use all the unit mesh as output terminals. Thus, whereas spin waves reflected from the outer frame with $\alpha$ = 1 are neglected in the previous papers, they affect the resultant spin waves in this study. 

The physical parameters used for the simulation are basically the same as those in the previous paper~\cite{Nakane2018IEEEAccess:_Reser_Compu_with_spin_waves_Excit_in_a_Garne_film}: the garnet film material is ${\mathrm{Tm_{3}Fe_{5}O_{12}}}$ or ${\mathrm{Y_{3}Fe_{5}O_{12}}}$, the saturation magnetization is $M_\textup{S} = 100$~kA/m, the exchange stiffness constant is $A_{\rm{EX}}=3.6\times10^{-12}$~J/m,  $K^{\rm H}_{\rm U}$ = 10~kJ/${\rm m^{3}}$, $K^{\rm L}_{\rm{U}}$=1~kJ/${\rm m^{3}}$, the cubic magnetic anisotropy constant is $K_\textup{C}$ = 0, the external magnetic field $H^{\rm{EX}}$ along the $z$-axis is 0.03~kA/m, the unit step time is $t_{0}=0.01~\rm{ns}$, and the simulation temperature is 0~K. By using the method in Ref. \cite{Baker2017}, we find that the ferromagnetic resonance (FMR) frequency is about 10~GHz when the entire garnet has $K^{\rm H}_{\rm U}$.

\subsection{Sinusoidal/square wave classification task}
\label{subsec:task}

We consider a task of sinusoidal/square wave classification in this paper. This is a standard task to evaluate the computational ability of a reservoir computing system in temporal pattern recognition. This task is different from the temporal exclusive OR (XOR) task in our previous paper\cite{ichimura2020a}. We modulate $K_{\rm{U}}(t)$ of time $t$ in the area of both two input electrodes 1 and 2 with sinusoidal or square waves having a fundamental frequency of $f=2.5~\rm{GHz}$ (the period $T_{0}=0.4$~ns). The sinusoidal waves are expressed as
\begin{eqnarray}
	K_{\rm{U}\rm{sin}}(t) = \frac{K^{\rm{H}}_{\rm{U}} + K^{\rm{L}}_{\rm{U}}}{2} + \frac{K^{\rm{H}}_{\rm{U}} - K^{\rm{L}}_{\rm{U}}}{2}\mathrm{cos}\biggl( \frac{2{\pi}t}{T_{0}}\biggr).
\label{eq:sin}
\end{eqnarray}
The square waves are approximated with a Fourier series to the 4th term as
\begin{eqnarray}
	K_{\rm{U}\rm{square}}(t) & = & \frac{K^{\rm{H}}_{\rm{U}} + K^{\rm{L}}_{\rm{U}}}{2} + \frac{K^{\rm{H}}_{\rm{U}} - K^{\rm{L}}_{\rm{U}}}{2} \biggl\{ \mathrm{cos}\biggl( \frac{2{\pi}t}{T_{0}}\biggr) 
\nonumber \\
& & -\frac{1}{3}\mathrm{cos}\biggl(\frac{3\times2{\pi}t}{T_{0}}\biggr)  +\frac{1}{5}\mathrm{cos}\biggl(\frac{5\times2{\pi}t}{T_{0}}\biggr) 
\nonumber \\
& & -\frac{1}{7}\mathrm{cos} \biggl(\frac{7\times2{\pi}t}{T_{0}}\biggr) \biggr\}.
\label{eq:term4}
\end{eqnarray}

\subsection{Readout processing and Learning}
\label{subsec:learning}

Fig.~\ref{fig:sond} shows also the total construction of the reservoir computing system including readout connections. The output of the reservoir is fed to a single neuron node. The size of the output electrode is identical with the mesh size in the present numerical analysis. The reservoir output is represented as the $x$-component of the spin vector $s_{x}(\bm{r}, n)$ at position $\bm{r} = ( x , y )$ at time step $n(=t/t_{0})$.
The reservoir output signals are transformed into a vector of envelope signals $\bm{x}(n)=[x({\bm{r}, n})] \in \mathbb{R}^{N_{\mathrm{o}} \times 1}$ \ ($N_{\mathrm{o}}$~: number of output electrodes) as
\begin{eqnarray}
\hspace*{-1em}
x(\bm{r}, n)
=
{\mathrm{Amp}} (s_{x}(\bm{r}, n)-s_{x}^{\mathrm{offset}}(\bm{r})) ,
\label{eq:yn} 
\end{eqnarray}
where ${\rm{Amp}}(\cdot)$ and $s_{x}^{\mathrm{offset}}(\bm{r})$ denote amplitude and direct current (dc) component, respectively. 
In this analysis, $s_{x}^{\mathrm{offset}}(\bm{r})$ is the initial value of $s_{x}$, i.e, $s_{x}(\bm{r}, 0)$. This amplitude extraction is realized by a diode and a low pass filter very easily in reality.

The envelope signals are weighted by the readout connection weights, denoted by $\mathbf{W}_{\mathrm{out}} \in \mathbb{R}^{1 \times N_{\mathrm{o}}}$, and then fed to the output neuron node. 
The output $\hat{\bm{y}}(n)$ at time step $n$ is given with an activation function $f$ working element-wise as
\begin{eqnarray}
\hat{\bm{y}}(n) = f( \mathbf{W}_{\rm{out}} \  {\bm{x}(n)} ) .
\label{eq:yhat}
\end{eqnarray}
Here we use a sigmoid function $f(u)=1/(1+e^{-u})$ in the following experiments.

In the learning of $\mathbf{W}_{\rm {out}}$, we use $ {\bf{X}} \equiv [\bm{x}(1)~\bm{x}(2)~...~\bm{x}(N)] \in \mathbb{R}^{N_{\mathrm{o}} \times N}$ as
\begin{eqnarray}
\mathbf{W}_{\mathrm{out}}=f^{-1}(\bf{Y})\bf{X}^{+} ,
\label{eq:W}
\end{eqnarray}
where $\bf{X}^{+}\in \mathbb{R}^{N \times N_{\mathrm{o}}}$ is the pseudo inverse matrix of $\bf{X}$, $N$ is the number of time steps used for the learning, ${\bf{Y}}$ is the matrix of teacher signals expressed as ${\bf{Y}} = [\bm{y}(1)~\bm{y}(2)~...~\bm{y}(N)] \in \mathbb{R}^{1 \times N}$, in which every component (only a single component $y(n)$ in $\bm{y}(n)$ for the single neuron here) is chosen as  
 \begin{eqnarray}
y(n) = \begin{cases} 
		0 & (\rm{in\ sin\mathchar`-wave\ time\ section})\\
		1 & (\rm{in\ square\mathchar`-wave\ time\ section})
	\end{cases} .
\label{eq:y}
\end{eqnarray}
%
To calculate the value of the inverse of the activation function, we used 0.001 and 0.999 instead of 0 and 1 as the teacher signals.

\begin{figure}
\centering
\includegraphics[width=0.2\hsize]{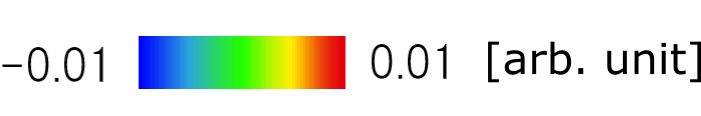} \\
\includegraphics[width=0.2\hsize]{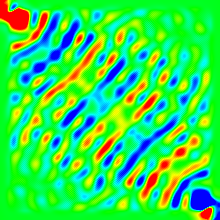} \ 
\includegraphics[width=0.2\hsize]{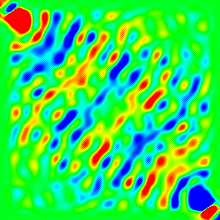} \\
0.000\ $T_{0}$ \hspace*{0.25\hsize} 0.025\ $T_{0}$  \vspace{5pt}\\
\includegraphics[width=0.2\hsize]{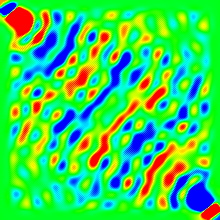} \ 
\includegraphics[width=0.2\hsize]{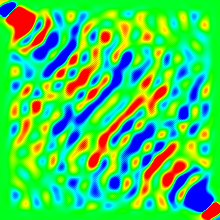} \\
0.050\ $T_{0}$ \hspace*{0.25\hsize} 0.075\ $T_{0}$  \vspace{5pt}\\
\includegraphics[width=0.2\hsize]{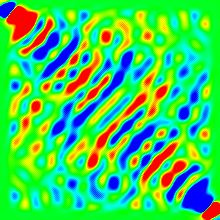} \ 
\includegraphics[width=0.2\hsize]{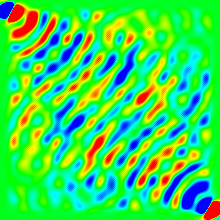} \\
0.100\ $T_{0}$ \hspace*{0.25\hsize} 0.125\ $T_{0}$  \vspace{5pt}\\
\includegraphics[width=0.2\hsize]{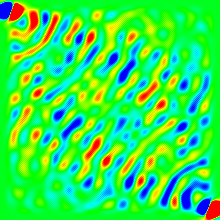} \ 
\includegraphics[width=0.2\hsize]{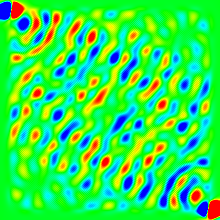} \\
0.150\ $T_{0}$ \hspace*{0.25\hsize} 0.175\ $T_{0}$  \vspace{5pt}\\
\includegraphics[width=0.2\hsize]{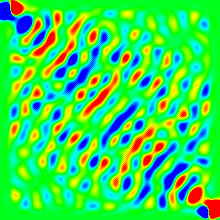} \ 
\includegraphics[width=0.2\hsize]{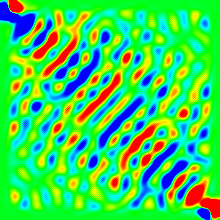} \\
0.200\ $T_{0}$ \hspace*{0.25\hsize} 0.225\ $T_{0}$  \vspace{5pt}\\
\caption{
Spin-wave propagation showing $x$-component spin $s_{x}$ when we feed the sinusoidal wave to the two input electrodes (period $T_{0}$=0.4~ns).
}
\label{fig:waves1xsin}
\end{figure}

\begin{figure}
\centering
\includegraphics[width=0.2\hsize]{./figures/colorbar_001.png} \\
\includegraphics[width=0.2\hsize]{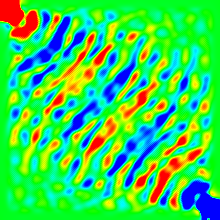} \ 
\includegraphics[width=0.2\hsize]{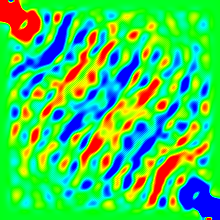} \\
0.000\ $T_{0}$ \hspace*{0.25\hsize} 0.025\ $T_{0}$  \vspace{5pt}\\
\includegraphics[width=0.2\hsize]{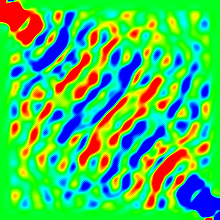} \ 
\includegraphics[width=0.2\hsize]{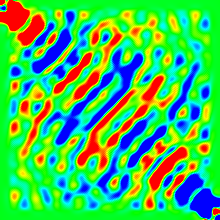} \\
0.050\ $T_{0}$ \hspace*{0.25\hsize} 0.075\ $T_{0}$  \vspace{5pt}\\
\includegraphics[width=0.2\hsize]{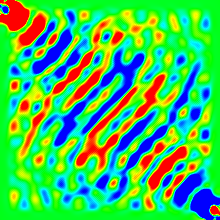} \ 
\includegraphics[width=0.2\hsize]{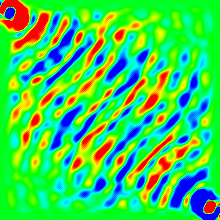} \\
0.100\ $T_{0}$ \hspace*{0.25\hsize} 0.125\ $T_{0}$  \vspace{5pt}\\
\includegraphics[width=0.2\hsize]{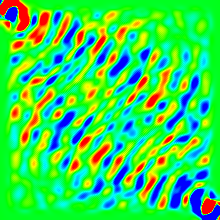} \ 
\includegraphics[width=0.2\hsize]{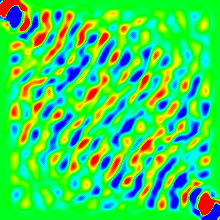} \\
0.150\ $T_{0}$ \hspace*{0.25\hsize} 0.175\ $T_{0}$  \vspace{5pt}\\
\includegraphics[width=0.2\hsize]{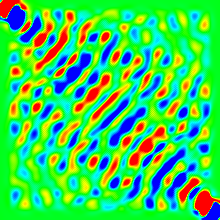} \ 
\includegraphics[width=0.2\hsize]{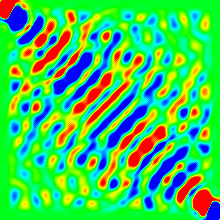} \\
0.200\ $T_{0}$ \hspace*{0.25\hsize} 0.225\ $T_{0}$  \vspace{5pt}\\
\caption{
Spin-wave propagation showing $x$-component spin $s_{x}$ when we feed the square wave to the two input electrodes (period $T_{0}$=0.4~ns).
}
\label{fig:waves4xsqr}
\end{figure}

\section{Numerical experiment and analyses}   
\label{sec:result}

\subsection{Spin-wave propagation}
\label{subsec:spinwave}

Fig.~\ref{fig:waves1xsin} shows the spin-wave propagation when we modulate $K_{\rm{U}}$ at the input electrodes with a sinusoidal wave having a frequency of $f=2.5~\rm{GHz}$, showing an almost steady state, from $t=0.000~T_{0}$ to $t=0.225~T_{0}$. We find that red and blue areas appear alternately in time almost everywhere in the spin-wave reservoir. Fig.~\ref{fig:waves4xsqr} shows the wave propagation when we feed a square wave. The interference of the spin waves in Fig.~\ref{fig:waves4xsqr} is similar to, but a little different from, that in Fig.~\ref{fig:waves1xsin}. The readout learns this difference. 

\begin{figure}
\centering
\includegraphics[width=0.25\hsize]{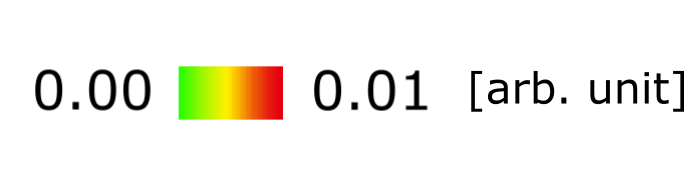} \\
\includegraphics[width=0.33\hsize]{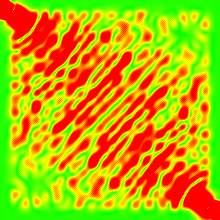} \\
(a)\\
\includegraphics[width=0.33\hsize]{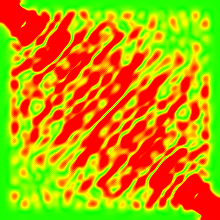} \\
(b)\\
\caption{Amplitude of the spin waves $\bm{x} ( n )$ at $t \ ( = n t _{0} ) = 0.000$~$T_{0}$ when we feed (a)sinusoidal and (b)square waves, respectively, to the two input electrodes.}
\label{fig:hil}
\end{figure}

Fig.~\ref{fig:hil} shows the distributions of the spin-wave amplitudes at 0.000~$T_{0}$ (input signal phase = 0) as an example. They are obtained by (\ref{eq:yn}) as the envelope, which is to be used as the input to the output neuron, $\bm{x}(n)$. Though the total distribution tendency is similar to each other, we can find differences in the details.

\subsection{Distribution of the readout weights}
\label{subsec:info}
\begin{figure}
\centering
\ \ \ \ 
\includegraphics[width=0.45\hsize]{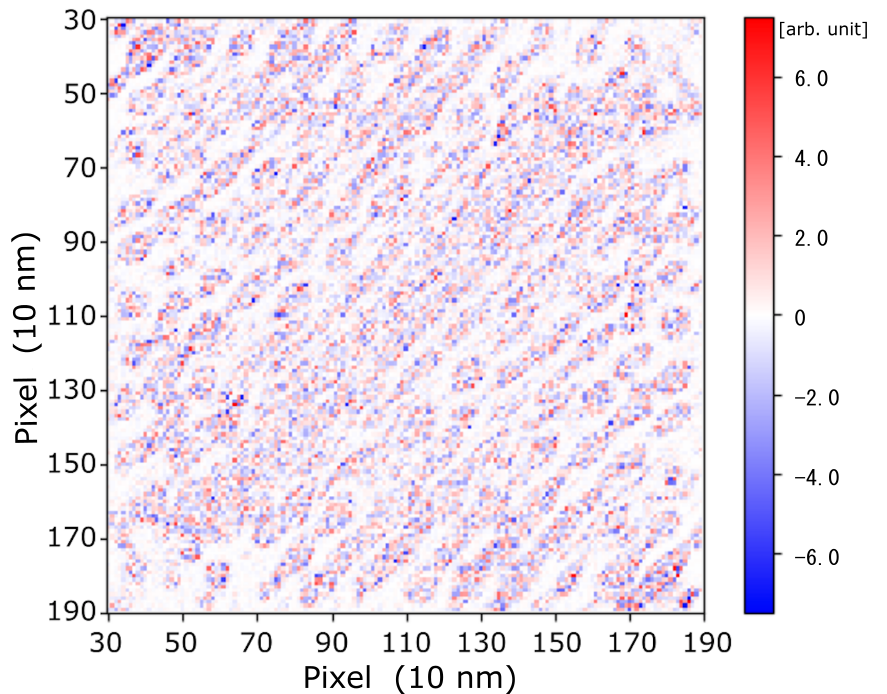}
\caption{Spatial distribution of the readout connection weights.}
\label{fig:weightsinsqr128_32i}
\end{figure}

In the following first analysis, every mesh unit (or pixel) in the center $160 \times 160$-pixel area, excluding damper and input electrode areas, has an output electrode, in order to investigate the distribution of effective information on the spin-wave reservoir chip. We train the output weights in the following way. We feed the input terminals with sin and square wave packets lasting for 1280 time steps (=$N$) sequentially. We call this time duration one section, and feed sin- or square-section signals time-sequentially, by choosing one of them at random, for 15 sections ($15\times1280$ time steps) in total.

Fig.~\ref{fig:weightsinsqr128_32i} shows the distribution of readout weight values after learning. A red point has a positive weight value while a blue point shows a negative one. More intense colors represent bigger absolute values.  
The texture of the output weights corresponds well to the wave-front amplitude in Fig.\ref{fig:hil}. This means that the learning process captures the features of the spin waves in this device appropriately.


\begin{figure}
\centering
\includegraphics[width=0.4\hsize]{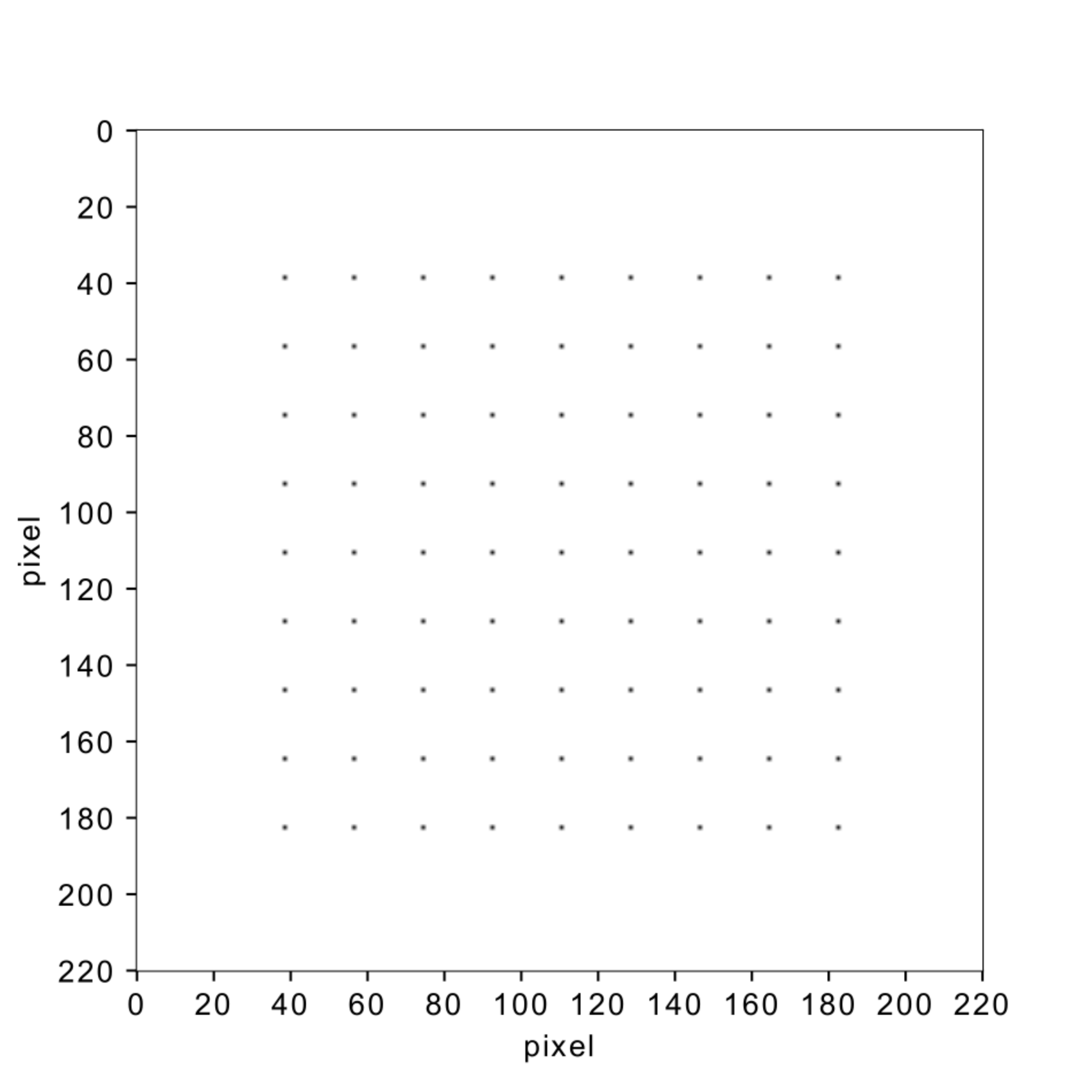} \\
(a) \\
\includegraphics[width=0.4\hsize]{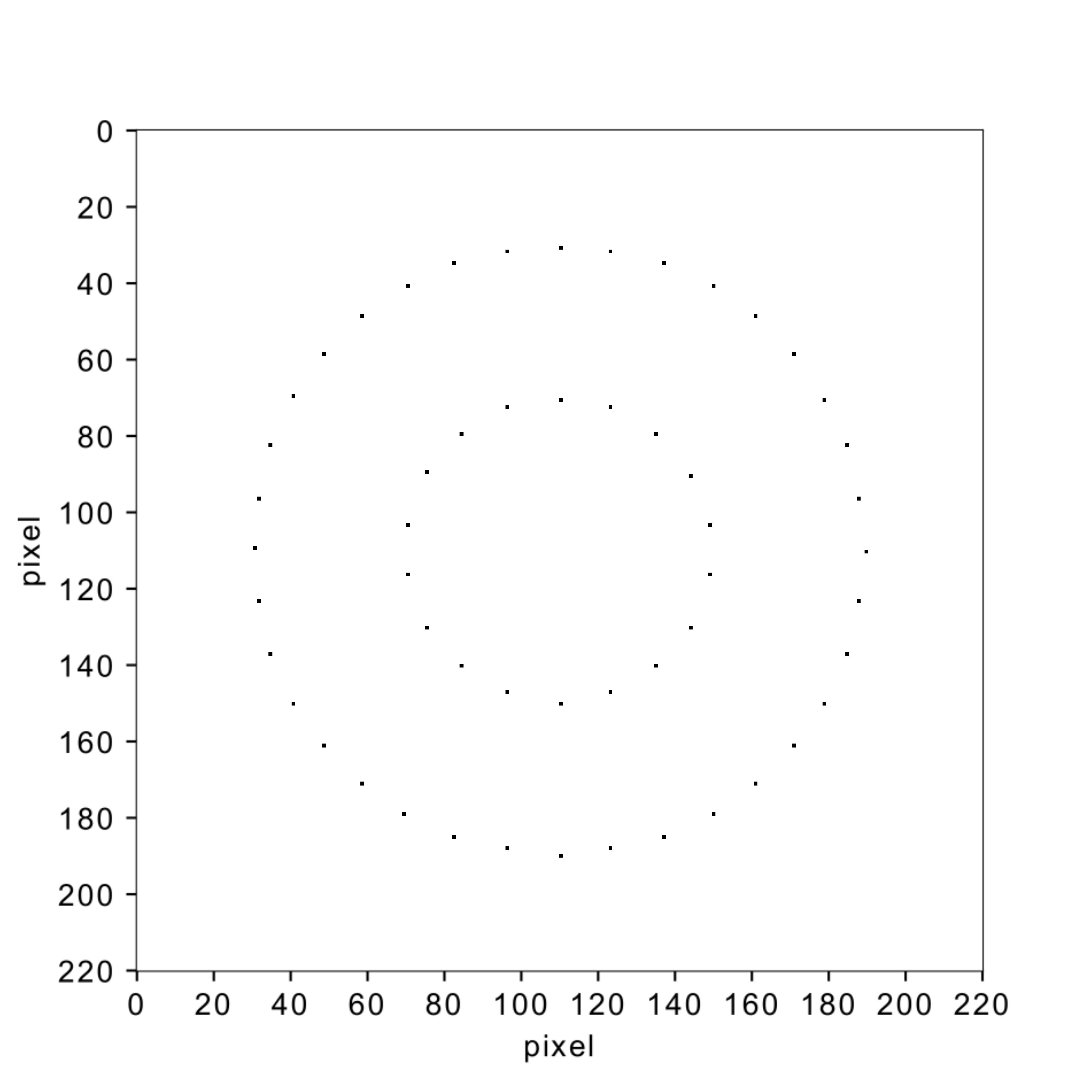} \\
(b) \\
\includegraphics[width=0.4\hsize]{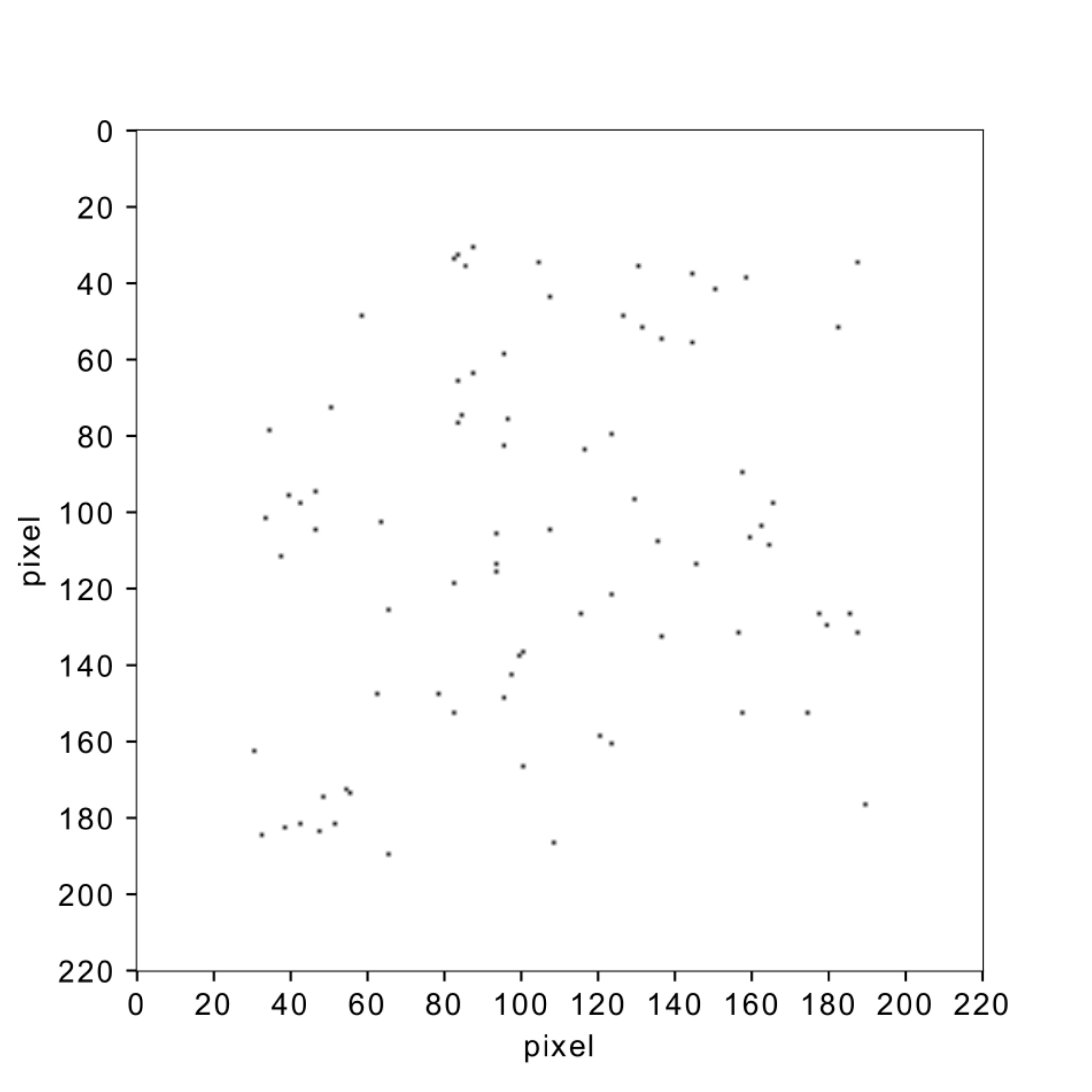} \\
(c)
\caption{Examples of the (a) grid, (b) circular and (c) random arrangements of the output electrodes when the number of output electrodes $N_{\mathrm{o}}$ is 81, 54 and 81, respectively.}
\label{fig:random}
\end{figure}

\subsection{Learning with a realistic number of output electrodes}
\label{subsec:limited}

In this section, we evaluate the learning and classification performance with a realistic number of output electrodes on the garnet film distributed at grid points, equiangularly circular points or at random for the same task. Fig.~\ref{fig:random} gives examples of the electrode positions for (a) grid  (the number of output electrodes is $N_{\mathrm{o}}$~=~81), (b) circular ($N_{\mathrm{o}}$~=~54), and (c) random ($N_{\mathrm{o}}$~=~81) arrangements, respectively. Since, in reality, it is impossible to put output electrodes on the damper area and on/around the input electrodes, we choose the central 160-pixel square area. In the grid case, we arranged the output electrodes with a $160/(\sqrt{N_{\mathrm{o}}} +1)$ pixel interval while, for the circular case, we put them on a single circle ($N_{\mathrm{o}}$~=~4, 16, 25) or double circles ($N_{\mathrm{o}}$~=~54, ...) with equal spacing. 


\begin{figure*}
\centering
\includegraphics[width=1\hsize]{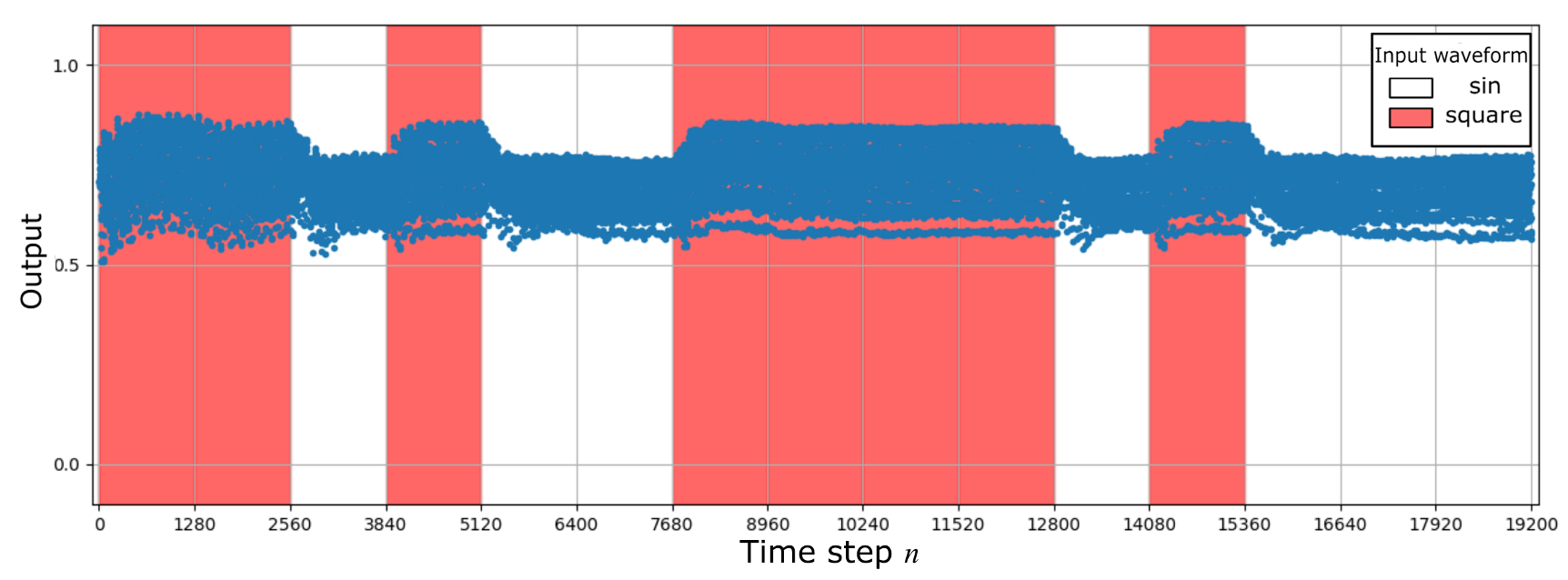} \\
(a)\\
\includegraphics[width=1\hsize]{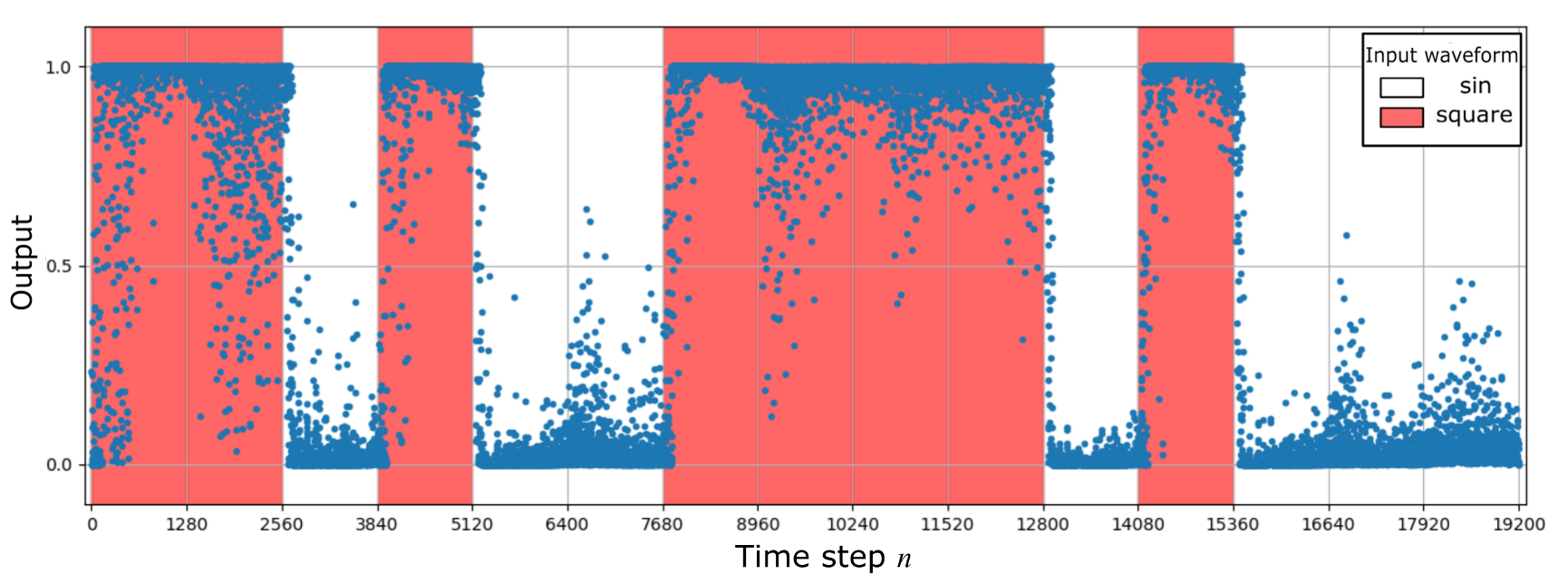} \\
(b)\\
\includegraphics[width=1\hsize]{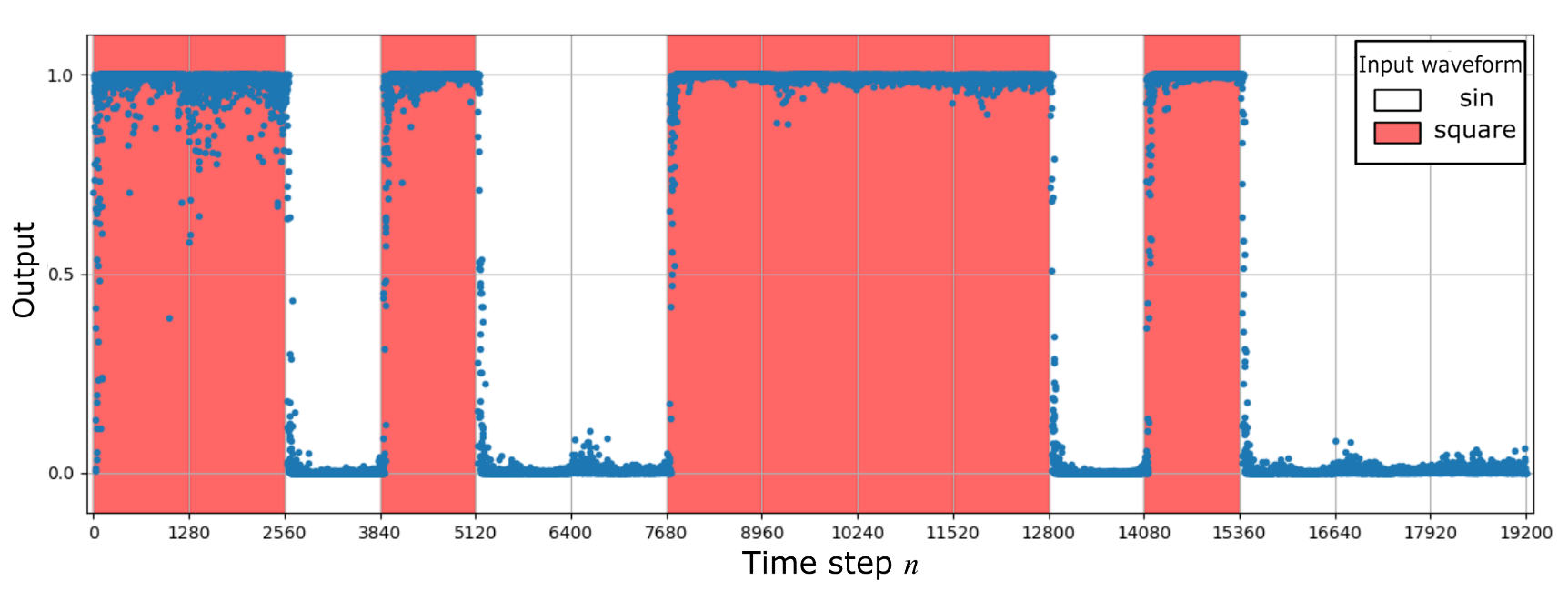} \\
(c)
\caption{Neural output signals for testing data for the number of output electrodes $N_{\mathrm{o}}$ of (a) 4, (b) 81 and (c) 289.}
\label{fig:batch}
\end{figure*}

Fig.~\ref{fig:batch} presents the corresponding neural output signals for (a) $N_{\mathrm{o}}$~=~32, (b) 81 and (c) 289 grid output-electrode cases.  They are the results for a testing section-series dataset that is different from the learning one. We find that, when a sinusoidal wave is input (white time section), the outputs are near to 0, while when a square wave is fed (red time section), the output is near to 1. We can also see that, as the number of the output electrodes increases, the classification becomes clearer.

The output neuron generates middle values in transient regions where the input signal is switched from sinusoidal to square or vice versa. This phenomenon corresponds to the fact that the reservoir deals with time-series data by utilizing its echoic nature. We can also find in Fig.\ref{fig:batch} that, as the number of the output electrodes increases, the response of the output against the sinusoidal/square wave switching becomes faster. This is because a larger number of output electrodes lead to a higher possibility to pick up fast responding signals such that the response is utilized for quick-switching realization.

\begin{figure}
\centering
\includegraphics[width=0.5\hsize]{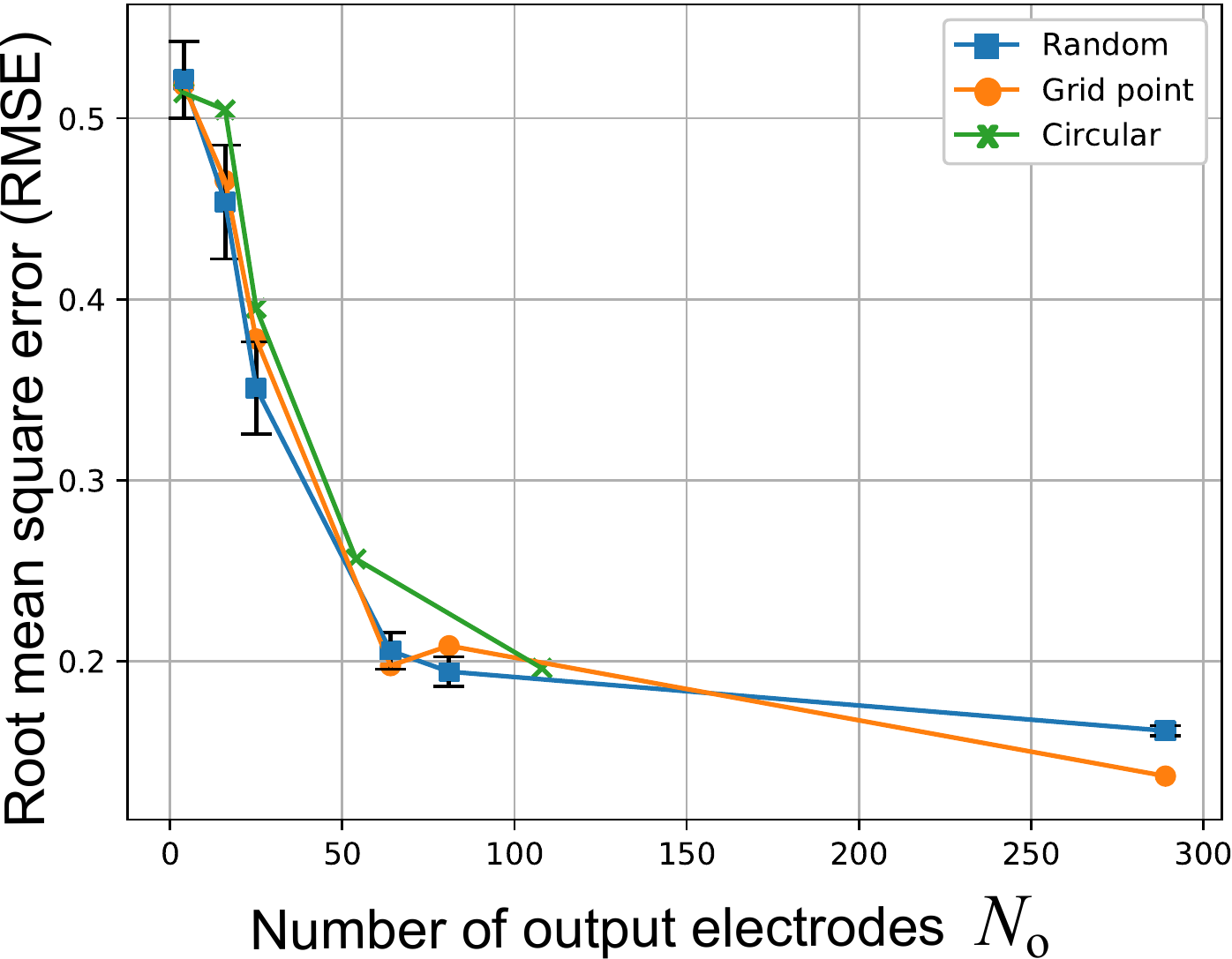}
\caption{Root mean square errors (RMSE) versus the number of output electrodes with error bars showing $\pm1\sigma$ (standard deviation).}
\label{fig:rmvs}
\end{figure}

\begin{figure}
\centering
\includegraphics[width=0.5\hsize]{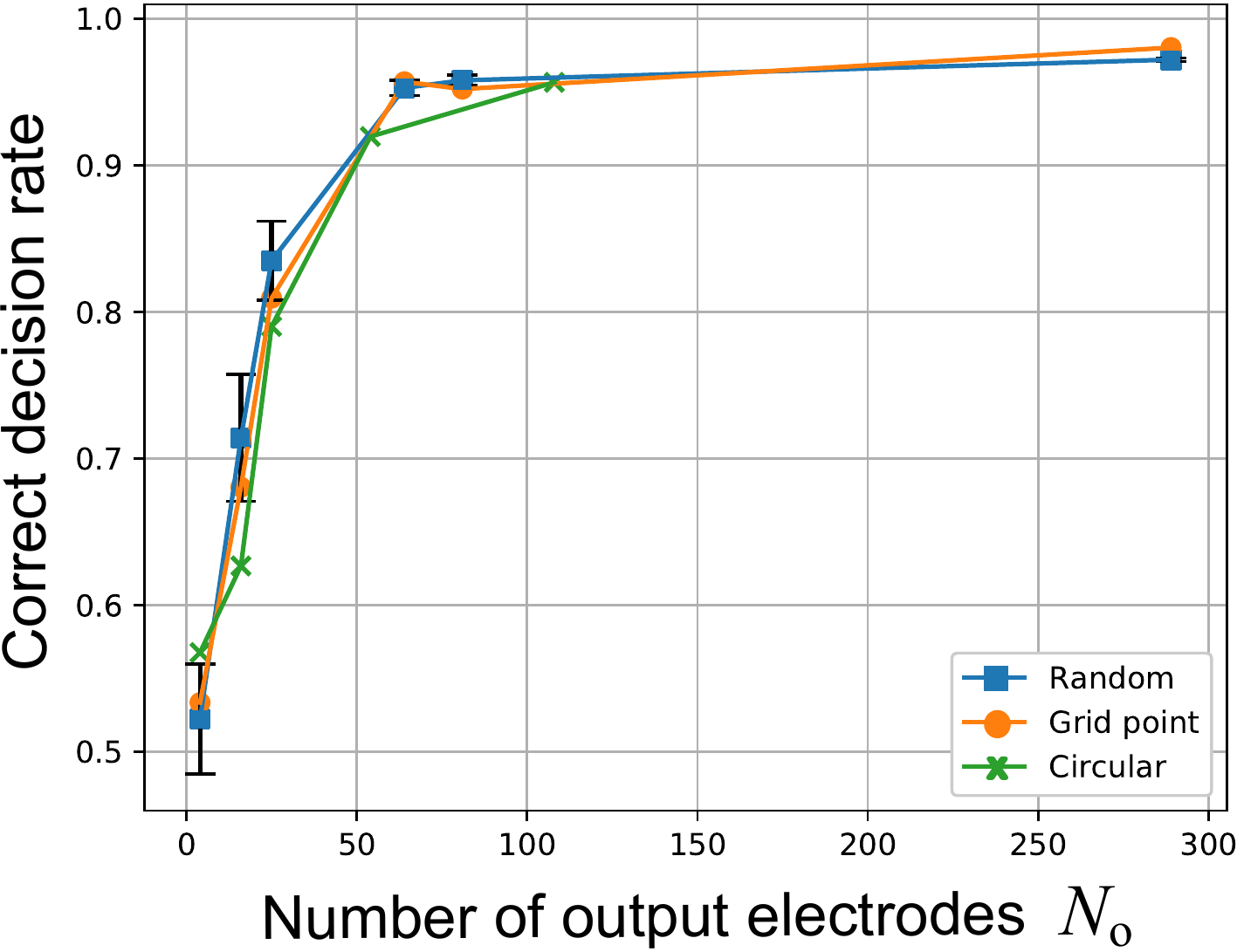} 
\caption{Correct decision rates versus the number of output electrodes with error bars showing $\pm1\sigma$ (standard deviation).}
\label{fig:crvs}
\end{figure}




Fig.~\ref{fig:rmvs} shows the root mean square errors (RMSE) versus the number of output electrodes, while Fig.~\ref{fig:crvs} presents the correct decision rates of classification on the test data when we choose the dichotomous threshold at 0.5. That is, the output is correct when $y(n)>0.5$ for the square wave and $y(n)\leq0.5$ for the sinusoidal wave. In this experiment, output electrode positions are selected at grid points, equiangularly circular points, or at random for ten times to evaluate the mean and the standard deviation. We find that the correct rates are over 90\% for about 50 electrodes or more. We can also find that the results are almost the same regardless of grid, circular or random arrangement. This fact indicates that the spin waves possess dynamics sufficiently diverse and asymmetric for this task.

\begin{figure}
\centering
\includegraphics[width=0.4\hsize]{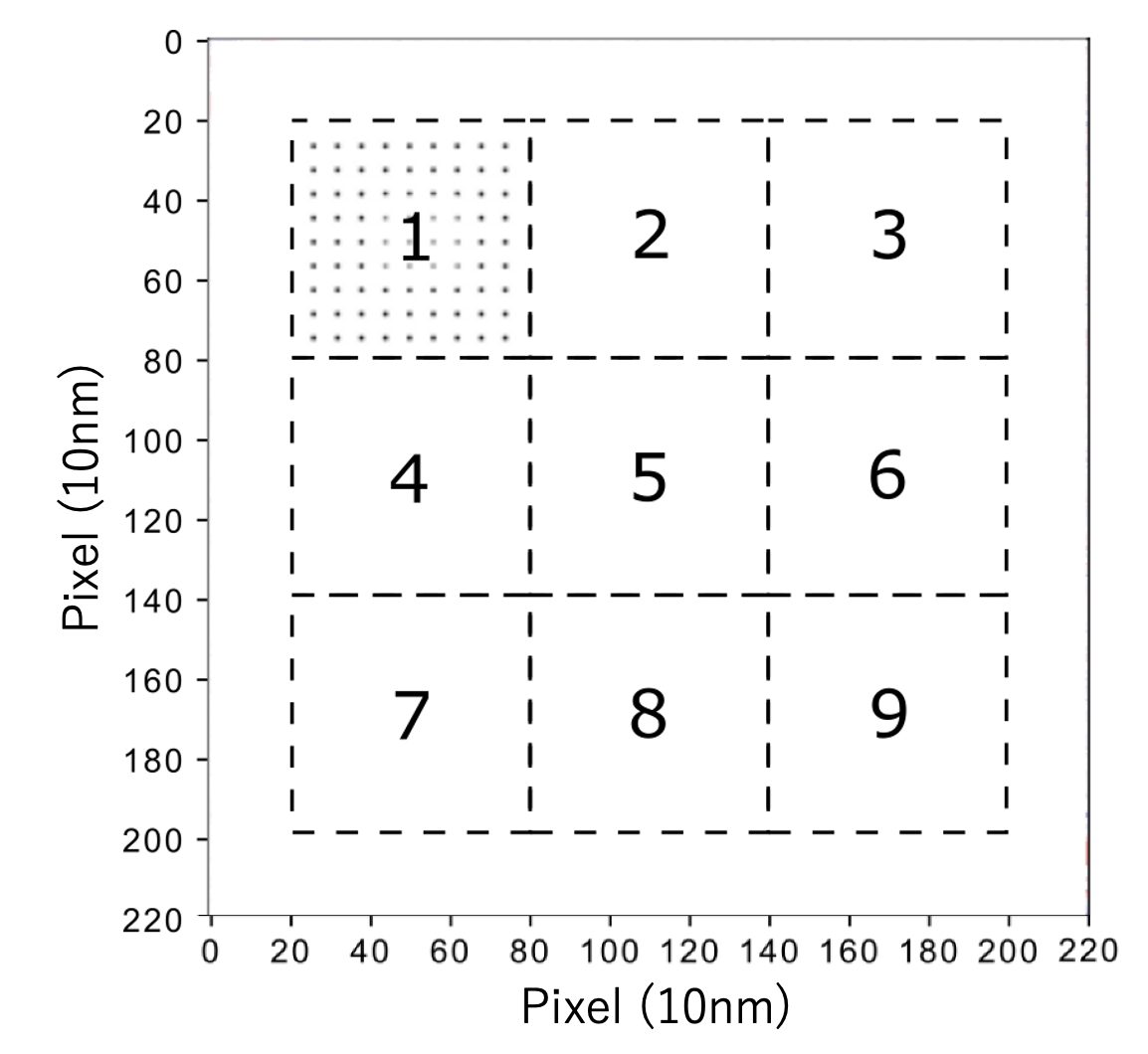} \\
(a) \\
\includegraphics[width=0.5\hsize]{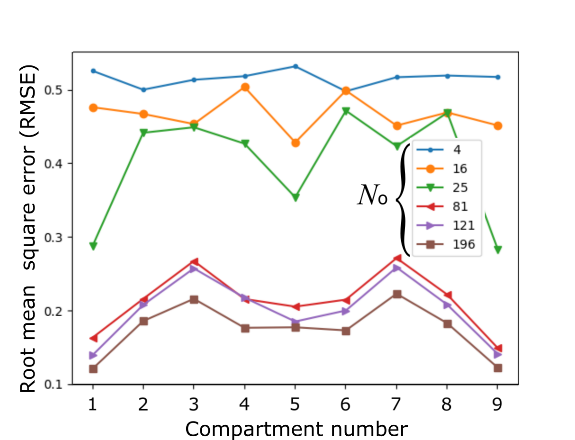} \\
(b) \\
\includegraphics[width=0.5\hsize]{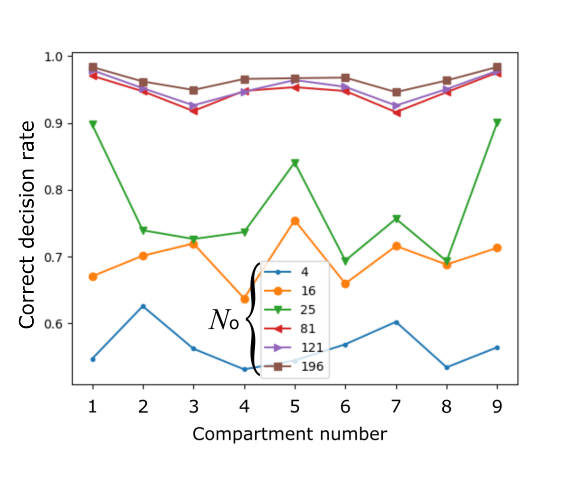} \\
\ \vspace*{-3.5em} \ \\
(c)
\caption{(a) Numbers assigned to respective compartments with a placement example of 81 electrodes in Compartment 1, (b) root mean square errors (RMSE) and (c)correct decision rates versus the compartment number when the output-electrode number $N_{\mathrm{o}}$ is varied from 4 to 196.}
\label{fig:place}
\end{figure}

\subsection{Useful-information distribution areas on the chip}
\label{subsec:place}

In this section, we investigate how the locations of output electrodes influence the performance. We divide the spin-wave propagation area of central $180 \times 180$ pixels into 9 compartments ($60 \times 60$ pixels) to place output electrodes in one of the compartments.


Fig.~\ref{fig:place}(a) shows the numbers assigned to respective compartments as well as 81 output-electrode positions at grid points as an example. Figs.~\ref{fig:place} (b) and (c) present RMSE and correct rates, respectively, versus the compartment number for various numbers of electrodes, $N_{\mathrm{o}}$. Fig.~\ref{fig:place} (b) shows that, when the number of the output electrodes $N_{\mathrm{o}}$ is small (4 - 16 electrodes), their placement is not so influential to RMSE. When $N_{\mathrm{o}}$ is 25, RMSE is decreased in Compartment 1 (near input electrode 1), 5 (at the center of the chip) and 9 (near input electrode 2). When $N_{\mathrm{o}}$~=~81 or over, the RMSE decreases and is particularly small for Compartments 1 and 9. In Fig.~\ref{fig:place} (c), when the number of electrodes $N_{\mathrm{o}}$ is small (4 or 16), the correct rate is low generally. When the number is 25, the correct rate is high in Compartment 1, 5 or 9. When $N_{\mathrm{o}}$ is 81 or larger, the dependence on the compartment is small and the correct decision rate is high in general.




The above dependence can be interpreted qualitatively as follows. In Fig.~\ref{fig:weightsinsqr128_32i}, we can see that more big absolute-value weights gather near the input electrodes. This suggests good results with output electrodes located in the compartments near the input electrodes. We can also see in Figs.~\ref {fig:waves1xsin} and \ref {fig:waves4xsqr} that there exist about six wave crests from the input electrodes to the center. Note that the input signal period is 0.4 ns, corresponding to 0.4ns / 0.01ns = 40 time steps. Then, it is possible that a 40$\times$6~=~240 time-step delay affects the switching performance in relation to the time needed for the spin waves to propagate to around the center of the chip. Actually, the output plotted in Figs.~\ref{fig:batch} (b) and (c) for total output electrodes $N_{\mathrm{o}}$ of 81 and 289 show that most errors occur at the transient time to switch from sin to square and vice versa with 200-300 time steps. These facts lead to the advantage of employing Compartment 1 or 9. The performance for Compartment 5 (center), slightly better than that of peripheral compartments except for 1 and 9, may be attributed to the large amplitude (Fig.~\ref{fig:hil}), like Compartments 1 and 9, which can enhance the sinusoid/square difference with nonlinearity.




\begin{figure}
\centering
\includegraphics[width=0.7\hsize]{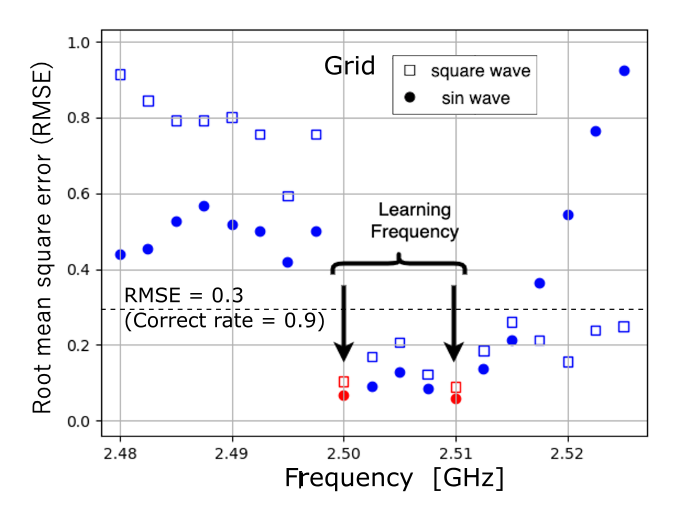}
(a)
\includegraphics[width=0.7\hsize]{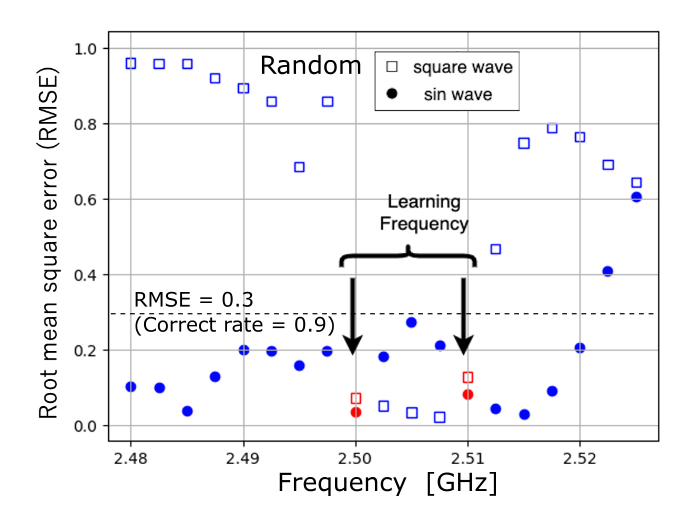}
(b)
\caption{RMSE versus frequency around 2.5~GHz for square ($\Box$) and sinusoidal ($\bullet$) waves with (a)~grid and (b)~random arrangements.}
\label{fig:general}
\end{figure}

\subsection{Generalization ability in the frequency domain}
\label{subsec:general}


We investigate the generalization ability of this device for the deviation of the fundamental signal frequency $f = 1/T_{0}$ in the sinusoidal/square wave classification task. A physical reservoir computing device using wave phenomena may show lower robustness against the changes of used frequency because, for example, pseudo-standing wave node position can be dependent on the frequency of the wave. In the following experiment, we check how the device is sensitive to such influence. 

Fig.~\ref{fig:general} shows the RMSE when the fundamental frequency of the learning and testing input signal is slightly changed around 2.5GHz.We conduct learning at two frequency points to examine the performance between these frequencies as well as its outside. Simulation temporal frame is 1280 time-step long for each frequency and waveform.  
Fig.~\ref{fig:general} is the RMSE for $N_{\mathrm{o}}$~=~81 with (a) grid  and (b) random arrangements, respectively. The output weights are obtained by using the pseudo inverse matrix as in (\ref{eq:W}). The red plots with arrows show the RMSE values at two learning frequency points while blue plots present those at other frequency points. We find that the RMSE values even at frequencies not used in the learning are still low when the frequency is in between or outside but the deviation is not large. The broken horizontal lines showing RMSE of 0.3 indicate a correct decision rate of about 0.9 (see Figs.~\ref{fig:place}(b) and (c)). This means that the device has generalization ability in the frequency domain, which may be favorable for applications with possible frequency deviations or certain bandwidth such as voice/sound treatment.

In addition, the grid result shown in Fig.~\ref{fig:general}(a) is roughly similar to the random results in Fig.~\ref{fig:general}(b), which corresponds well to Figs.~\ref{fig:rmvs} and \ref{fig:crvs}. This similarity suggests that the spin waves hold physical dynamics sufficiently complex and useful in the reservoir computing framework for this kind of tasks.

\section{Conclusion}
\label{sec:conclusion}

This paper numerically studied the influence of the arrangement of output electrodes on the spin-wave reservoir-computing performance. They are significantly important in designing a spin-wave reservoir chip since the spin waves transform input information into high-dimensional information space with their spatiotemporal dynamics, which determines reservoir-computational functionality. First, we observed the spin-wave propagation states. The task is sinusoidal/square wave classification, which is a basic one in time-series processing. We visualized the spatial distribution of useful information by plotting the weight values, which was found well reflecting the wave-front texture. Then, we investigated the influence of the output terminal number and positions on the performance.
We found that the classification was successfully performed with only several tens of electrodes, which is a reasonable number for fabrication.
%
This result suggested that the spin waves possess sufficiently complex and rich dynamics. This sufficiency was also indicated by the performance very similar for the grid, circular and random arrangements of the output electrodes. We also investigated in which area useful information is distributed more. We discussed that the transient performance can be influenced by the distance between the input and output electrodes. In addition, we confirmed that this device has generalization ability in the frequency domain. These results in total will lead to the establishment of practical design procedure of spin-wave reservoir devices for near-future low-power intelligent computing.


\bibliographystyle{IEEEtran}
\bibliography{mybiblio_access_GT_RN,reference_shimomura}

\end{document}